\documentclass[12pt]{article}
\usepackage{graphicx}

\newcommand{\HRule}{\rule{0.4\linewidth}{0.3mm}}

\newcommand{\jpsi}{J/$\psi$}
\newcommand{\psip}{$\psi^\prime$}
\newcommand{\jpsidy}{J/$\psi$\,/\,DY}
\newcommand{\chic}{$\chi_c$}

\newcommand{\ccbar}{$\rm c\overline{\rm c}$}
\newcommand{\bbbar}{$\rm b\overline{\rm b}$}

\newcommand{\pt}{$p_{\rm T}$}
\newcommand{\xf}{$x_{\rm F}$}
\newcommand{\EZDC}{$E_{\rm ZDC}$}
\newcommand{\ET}{$E_{\rm T}$}
\newcommand{\npart}{$N_{\rm part}$}
\newcommand{\sabs}{$\sigma_{\rm abs}$}
\newcommand{\sqrts}{$\sqrt{s}$}
\newcommand{\aptsquared}{$\langle p_{\rm T}^2 \rangle$}

\begin{document}

\begingroup
\thispagestyle{empty}
\baselineskip=14pt
\parskip 0pt plus 5pt

\begin{center}
{\large EUROPEAN LABORATORY FOR PARTICLE PHYSICS}
\end{center}

\bigskip
\begin{flushright}
October 4, 2006
\end{flushright}

\bigskip\bigskip\bigskip
\begin{center}
{\Large\bf
Open questions in quarkonium\\ [0.3cm]
and electromagnetic probes}

\bigskip\bigskip\bigskip
Carlos Louren\c{c}o\\ [0.2cm]
CERN-PH, CH-1211 Geneva 23, Switzerland

\bigskip\bigskip\bigskip\bigskip\bigskip\bigskip
\textbf{Abstract}
\end{center}

\begingroup
\leftskip=0.4cm
\rightskip=0.4cm
\parindent=0.pt

In my (``not a summary'') talk at the Hard Probes 2006 conference, I
gave ``a personal and surely biased view on only a few of the many
open questions on quarkonium and electromagnetic probes''.  Some of
the points reported in that talk are exposed in this paper, having in
mind the most important of all the open questions: do we have, today,
from experimental data on electromagnetic probes and quarkonium
production, convincing evidence that shows, beyond reasonable doubt,
the existence of ``new physics'' in high-energy heavy-ion collisions?

\endgroup

\vfill
\noindent
\HRule

\noindent
Invited talk at the 2nd International Conference on Hard and
Electromagnetic Probes of High-Energy Nuclear Collisions, Asilomar,
California, June 9--16, 2006.
To be published in Nuclear Physics A.

\endgroup

\newpage


The first experimental measurements of photon and dilepton production
in high-energy heavy-ion collisions took place 20 years ago, at the
CERN SPS, with Oxygen ions accelerated to 200~GeV per nucleon.  Many
difficulties make these measurements particularly challenging.  In
particular, the production cross-sections of such particles as the
\jpsi\ and \psip\ are very low, the branching ratios of the decay
channels $\rho, \omega, \phi\to l^+l^-$ are very small, the
backgrounds from $\pi^0\to \gamma\gamma$ and other hadronic decays are
very high, etc.  Under these conditions, it is quite remarkable that
significant physics progress actually resulted from the first
generation of SPS experiments (none of the AGS heavy-ion experiments
measured dileptons or photons).  It is important to keep in mind that
these experiments were built in a very short time and, most of them if
not all, used existing detectors with only minimal additions or
modifications.  For instance, the NA38 experiment was proposed to the
SPS committee in 1985 (to search for thermal dimuons) and was taking
data one year later, with newly built beam detectors, an active target
system and an electromagnetic calorimeter, added to the muon
spectrometer used by NA10 since 1980.  Now that most people in our
field work in experiments that cost in excess of 100 million Swiss
francs, it might be worth recalling that the (much cheaper) 17~m long muon
spectrometer built by NA10 worked in the NA10 experiment for 5 years
and in the NA38, NA50 and NA60 ``heavy-ion experiments'' for 18 years!
Surely a very good deal.

Already in this first generation of SPS heavy-ion experiments, using
beams of Oxygen and Sulphur ions, several observations indicated the
presence of ``anomalies'' in the measured data.  Until 1995, many
people argued that the drop of the ratio between the \jpsi\ and the
``Drell-Yan continuum'' production cross-sections, from p-U to O-U to
S-U collisions and from peripheral to central O-U and S-U collisions,
as measured by NA38, constituted a clear signature of QGP formation,
following the prediction published in 1986, in what is nowadays
arguably the most famous paper of our field~\cite{MS1986} (cited
almost 1000 times by now).

In parallel, an excess in the production of low mass dielectrons was
found by the CERES collaboration and interpreted as a possible
indication that chiral symmetry was (approximately) restored in the
matter produced in S-Au collisions at SPS energies.

Also the continuum dimuons with masses below the \jpsi\ peak were
under scrutiny, by NA38 and Helios-3.  Both experiments found an
excess in the production yield of such dimuons in Sulphur induced
collisions, with respect to the yields expected from a superposition
of Drell-Yan dimuons and muon pairs from simultaneous semi-muonic
decays of (correlated) pairs of D mesons.  These observations were
taken seriously because the same analyses could reproduce the data
collected in proton-nucleus collisions, with normalisations compatible
with the expectations available at the time.  Possible explanations of
this ``intermediate mass region dimuon excess'' included thermal
dimuons (maybe emitted from the QGP phase) and modifications of the
pair correlations of the D mesons while traversing the medium produced
in the nuclear collisions.

\begin{figure}[ht]
\centering
\resizebox{0.48\textwidth}{!}{%
\includegraphics*{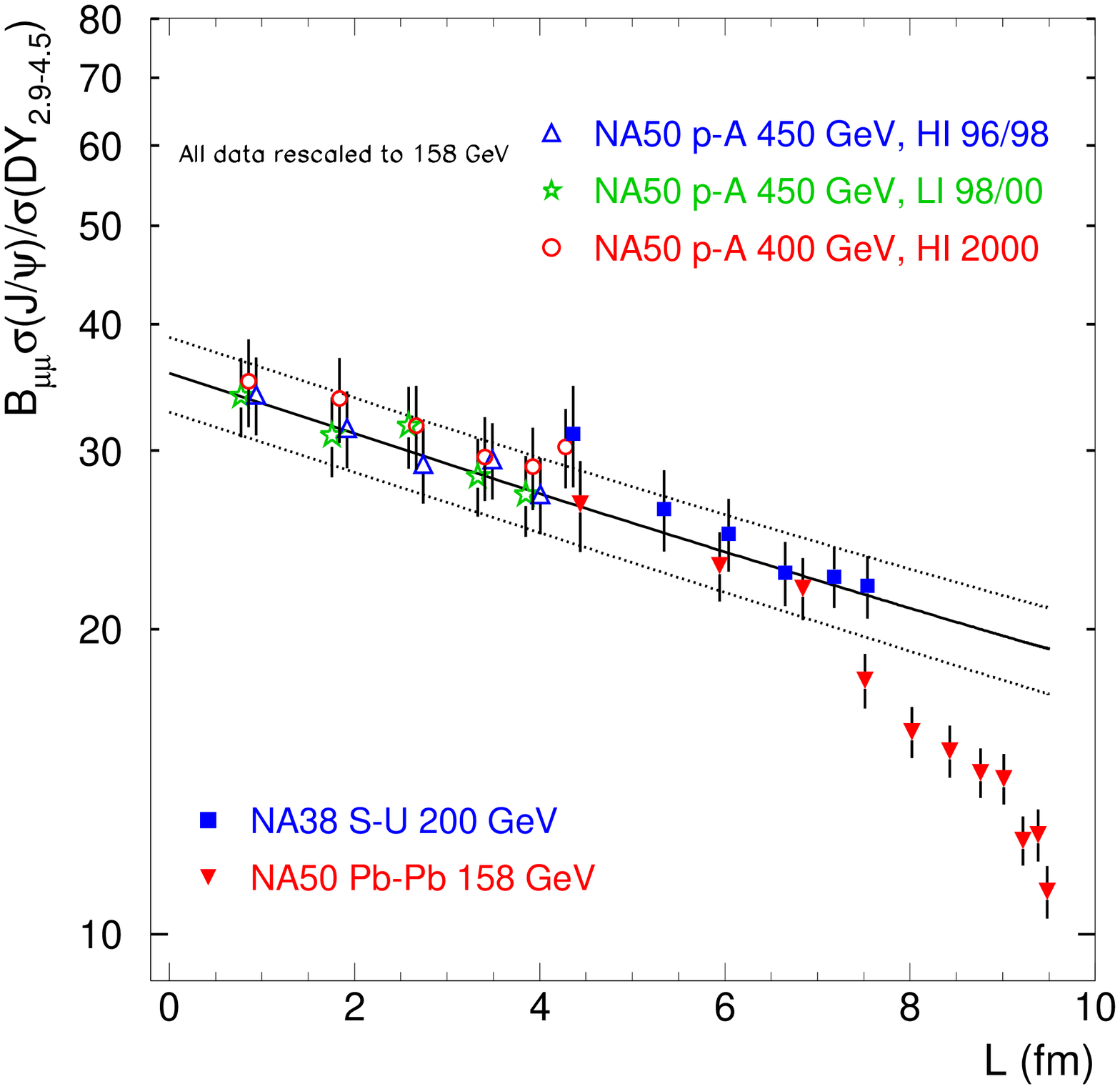}}
\resizebox{0.48\textwidth}{!}{%
\includegraphics*{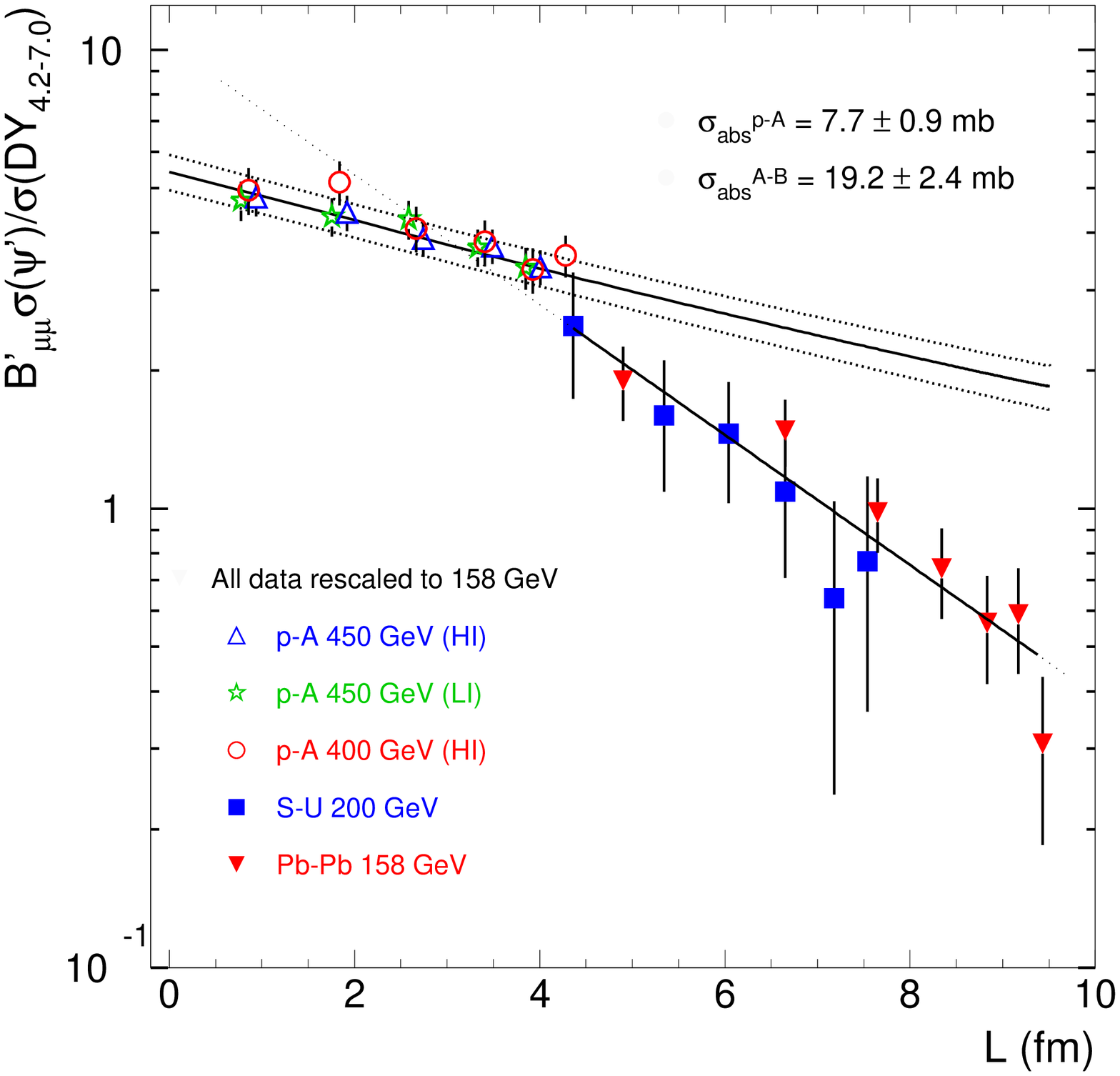}}
\caption{Left: The \jpsi\ suppression pattern as measured by NA38 and
  NA50, in p-A, S-U and Pb-Pb collisions, as a function of L, the
  distance of nuclear matter traversed by the charmonium, compared to
  the ``normal nuclear absorption'' band resulting from the Glauber
  fit to the p-A data~\cite{Goncalo}.  The p-A and S-U data points, as
  well as the curves, have been rescaled to the conditions (energy and
  rapidity window) of the Pb-Pb data.  Right: The suppression of the
  \psip\ as measured by NA38 and NA50 in p-A, S-U and Pb-Pb
  collisions~\cite{Helena}, compared to the normal nuclear absorption
  curve calculated using the Glauber formalism~\cite{KLNS}.}
\label{fig:na38-na50}
\end{figure}

Between 1994 and 2000, a second important step took place at the SPS,
with several data taking periods with Pb ions, at 158~GeV and lower
energies.  The new experiments confirmed the observation of excess
production of low and intermediate mass dileptons, now in Pb induced
collisions.  The situation changed, however, in the case of the
production and suppression of charmonium states, in particular thanks
to a much better understanding of the proton-nucleus reference data.
In the case of the \jpsi, we now think that the ``normal nuclear
absorption'' already present in p-A collisions, and nicely reproduced
by a simple model based on the Glauber formalism with a single
``absorption cross section'' parameter, \sabs, can also account for
the ``suppression pattern'' measured in S-U collisions, with the same
\sabs\ value, indicating that the \jpsi\ is not sensitive to the
medium formed in such collisions, contrary to what happens in Pb-Pb
collisions, where an extra and significant suppression is seen, beyond
the expected ``normal nuclear absorption curve'', for the most central
collisions (Fig.~\ref{fig:na38-na50}-left).  In the case of the \psip,
on the other hand, there is a very significant extra suppression
observed already in S-U collisions, with respect to the p-A reference,
and no significant change is visible between the S-U and Pb-Pb
suppression patterns (Fig.~\ref{fig:na38-na50}-right).

It can be easily argued that one of the most important ``discoveries''
of the first 10 years of heavy-ion experimentation at the CERN SPS,
and not only in what concerns hard and electromagnetic probes, is that
it is extremely important, indeed crucial, to have very robust data
collected in more elementary collision systems, such as proton-proton
and proton-nucleus interactions, collected by the same experiments, at
the same collision energies, and in the same phase space acceptance
windows, as the heavy-ion measurements.  Only after having a solid
``expected baseline'', provided by a good understanding of the
relevant physics processes and based, in particular, on detailed
analyses of p-A data, we can realistically hope to identify patterns
in the high-energy heavy-ion data that will clearly and convincingly
signal the presence of ``new physics'' in the matter produced in those
interactions.

It is important to recognise the effort made at RHIC in this respect,
with pp and d-Au runs that have provided extremely valuable
information to understand the observations made in Au-Au collisions.
Unfortunately, for certain physics topics, such as \jpsi\ production,
the presently available d-Au statistics is very limited; a
high-statistics d-Au run should take place at RHIC at the earliest
possible date.

\begin{figure}[ht]
\centering
\resizebox{0.48\textwidth}{!}{%
\includegraphics*{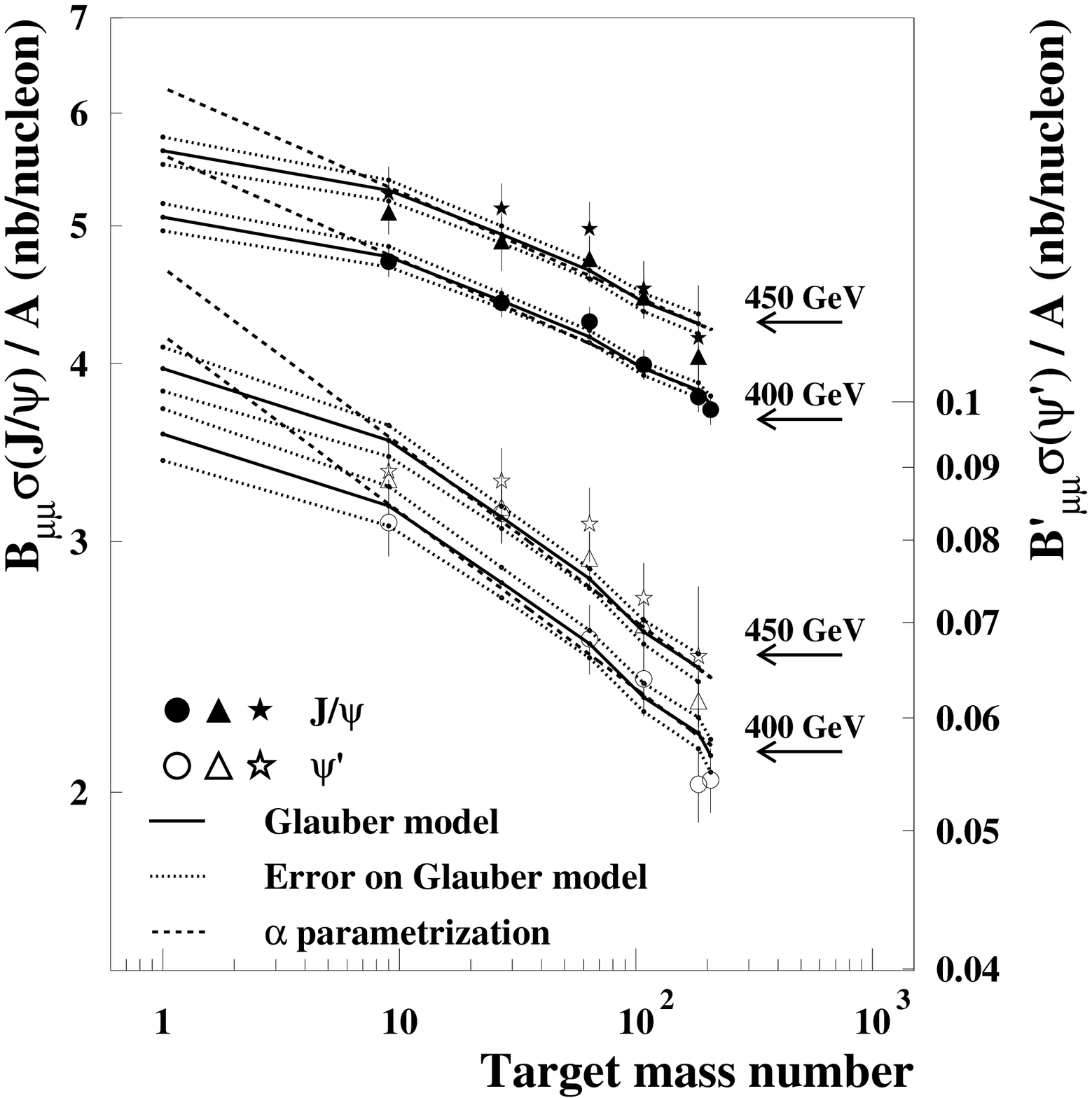}}
\resizebox{0.48\textwidth}{!}{%
\includegraphics*{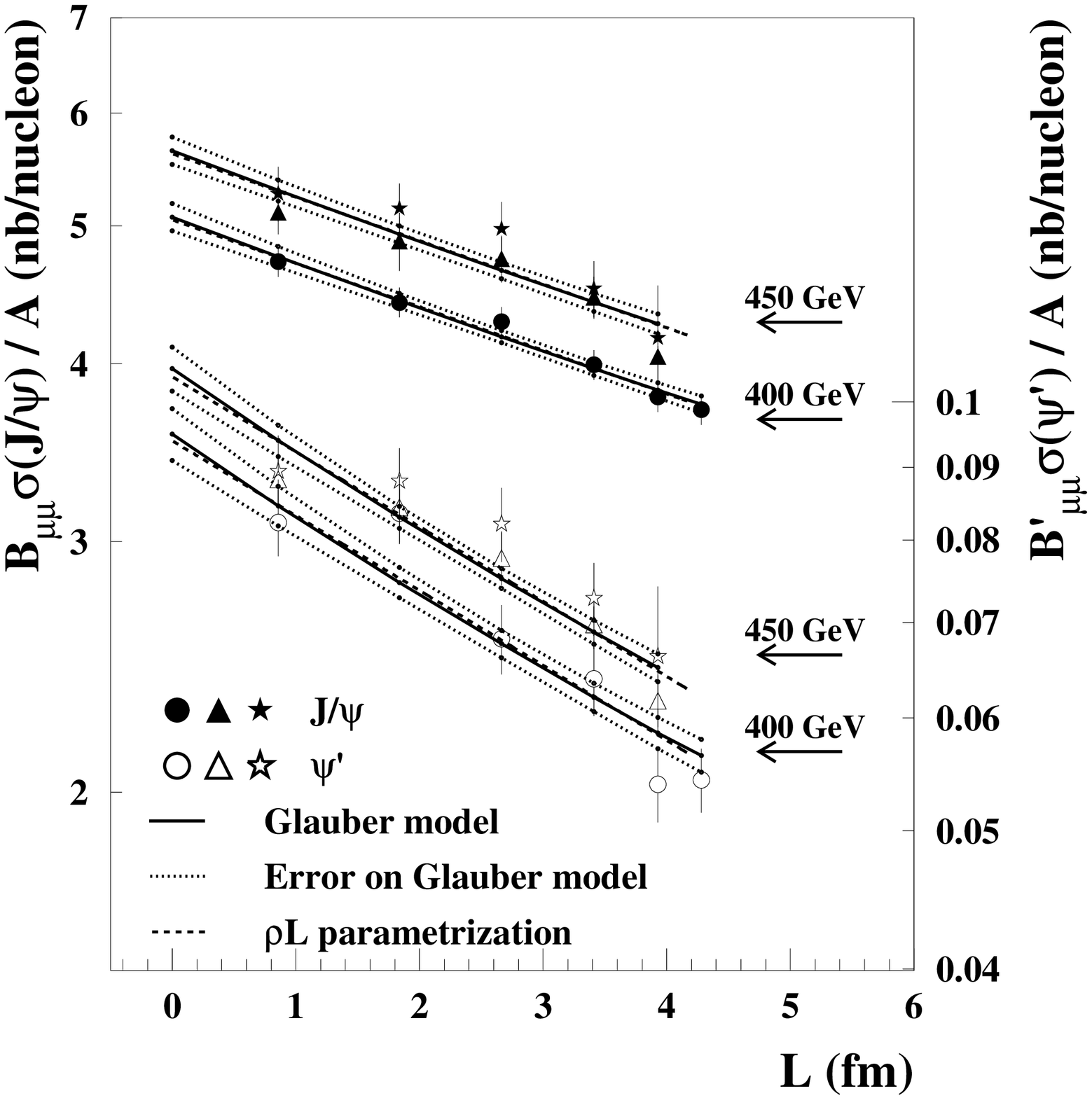}}
\caption{The \jpsi\ and \psip\ production cross sections measured in
  p-A collisions by NA50, as a function of the mass number of the
  target nuclei (left), and of L (right).  The lines represent fits to
  three commonly used parameterizations of the ``normal nuclear
  absorption''.}
\label{fig:NA50pA}
\end{figure}

At the SPS energies, the NA50 experiment has recently provided a very
detailed analysis of \jpsi\ and \psip\ production in p-A
collisions~\cite{Goncalo}, at 450 and 400~GeV, using 5 and 6 different
nuclear targets, respectively, both in terms of absolute production
cross sections (Fig.~\ref{fig:NA50pA}) and in terms of ratios between
charmonia and Drell-Yan yields (Fig.~\ref{fig:na38-na50}).

The results, corresponding to 16 different data sets collected between
1996 and 2000, in different experimental conditions, give a rather
consistent picture of the \jpsi\ and \psip\ ``normal nuclear
absorptions'', with absorption cross sections $\sigma_{\rm abs}^{{\rm
J}/\psi} = 4.5\pm0.5$~mb and $\sigma_{\rm abs}^{\psi^\prime} =
8.3\pm0.9$~mb from the production cross-sections, and $\sigma_{\rm
abs}^{{\rm J}/\psi} = 4.2\pm0.5$~mb and $\sigma_{\rm
abs}^{\psi^\prime} = 7.7\pm0.9$~mb from the cross-section ratios.

\begin{figure}[ht]
\centering
\resizebox{0.48\textwidth}{!}{%
\includegraphics*{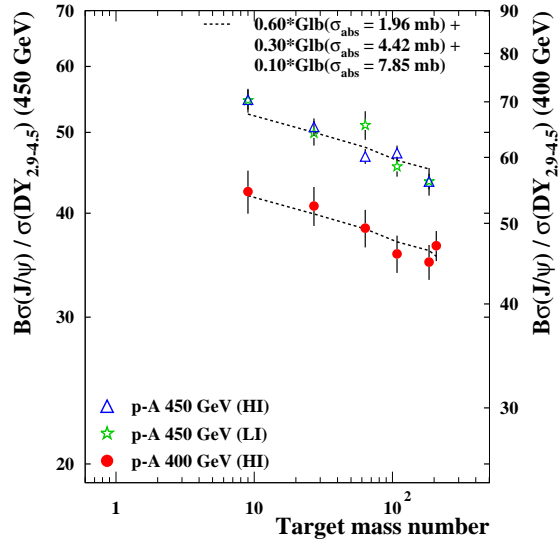}}
\caption{NA50 p-A values of the \jpsi\ over Drell-Yan
  cross-section ratio, as a function of A, compared to a Glauber
  calculation taking into account the feed-down contributions from
  higher \ccbar\ states.}
\label{fig:feeddown}
\end{figure}

However, it is important to keep in mind that a significant fraction
of the observed \jpsi\ mesons result from decays of \psip\ and \chic\
mesons, that the \psip\ has a higher \sabs\ value (at least at
mid-rapidity), and that this is likely to also be the case for the
\chic, a larger and more weakly bound charmonium state than the \jpsi.
If we assume that 60\,\% of the observed \jpsi\ mesons are directly
produced, 30\,\% result from \chic\ decays and 10\,\% from \psip\
decays, and if, furthermore, we calculate $\sigma_{\rm abs}$
geometrically, as $\pi r^2$, with $r_{{\rm J}/\psi}=0.25$~fm,
$r_{\psi^\prime}=2\times r_{{\rm J}/\psi}$ and $r_{\chi_c}=1.5\times
r_{{\rm J}/\psi}$, we can redo the Glauber calculation, now without
leaving \sabs\ as a free parameter, and we obtain an equally good
description of the \jpsi\ data (Fig.~\ref{fig:feeddown}), with a
reduced $\chi^2$ of 1.0 (by chance?).  Besides, the value of
$\sigma_{\rm abs}^{\psi^\prime}$ from this very simple model is
7.85~mb, in perfect agreement with the NA50 measurements (by chance?).

\begin{figure}[ht]
\centering
\resizebox{0.48\textwidth}{!}{%
\includegraphics*{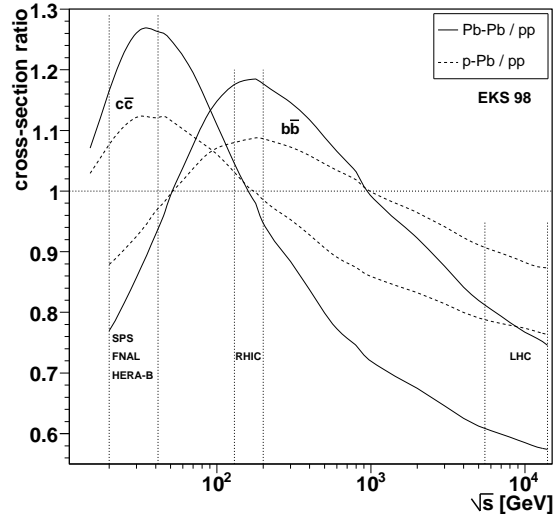}}
\caption{Changes induced on the \ccbar\ (and
  \bbbar) cross-sections, in p-Pb and Pb-Pb collisions, by the nuclear
  modifications of the PDFs, at mid-rapidity~\cite{PhysRep}.}
\label{fig:npdfs}
\end{figure}

The values of \sabs\ mentioned here are ``effective'' values,
corresponding to a convolution of the final state absorption which
affects the produced \ccbar\ states while traversing the nuclear
matter, with initial state effects, such as the nuclear effects on the
gluon distribution functions.  According to the well-known EKS98 model
of nuclear effects on the parton distribution functions~\cite{EKS98},
charm production at SPS energies, and at mid-rapidity, is in the
anti-shadowing region, leading to an initial state \ccbar\ production
enhanced in p-Pb collisions, say, with respect to the case when
nuclear effects are neglected (Fig.~\ref{fig:npdfs}).

A very simple calculation indicates that the ``convoluted'' \sabs\
value, 4.2~mb, becomes around 6~mb if we explicitly take into account
the anti-shadowing effect, using the EKS98 model.  This value is more
directly comparable to the values obtained from the d-Au data of
PHENIX, between 0 and 3~mb, which were also obtained after taking into
account the (EKS98) nuclear effects on the PDFs~\cite{PHENIX-psi}.

\begin{figure}[h!]
\centering
\resizebox{0.48\textwidth}{!}{%
\includegraphics*{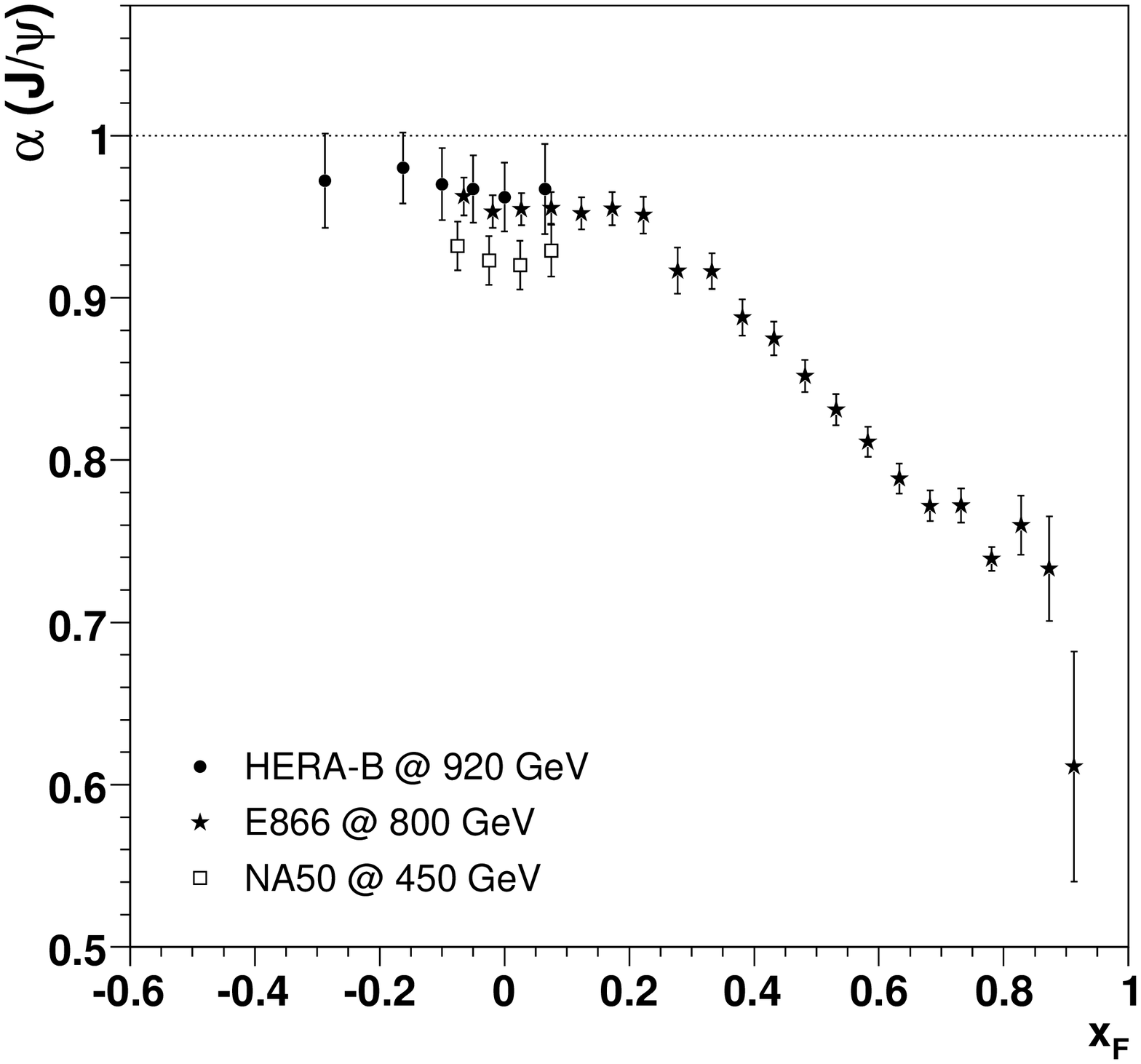}}
\resizebox{0.48\textwidth}{!}{%
\includegraphics*{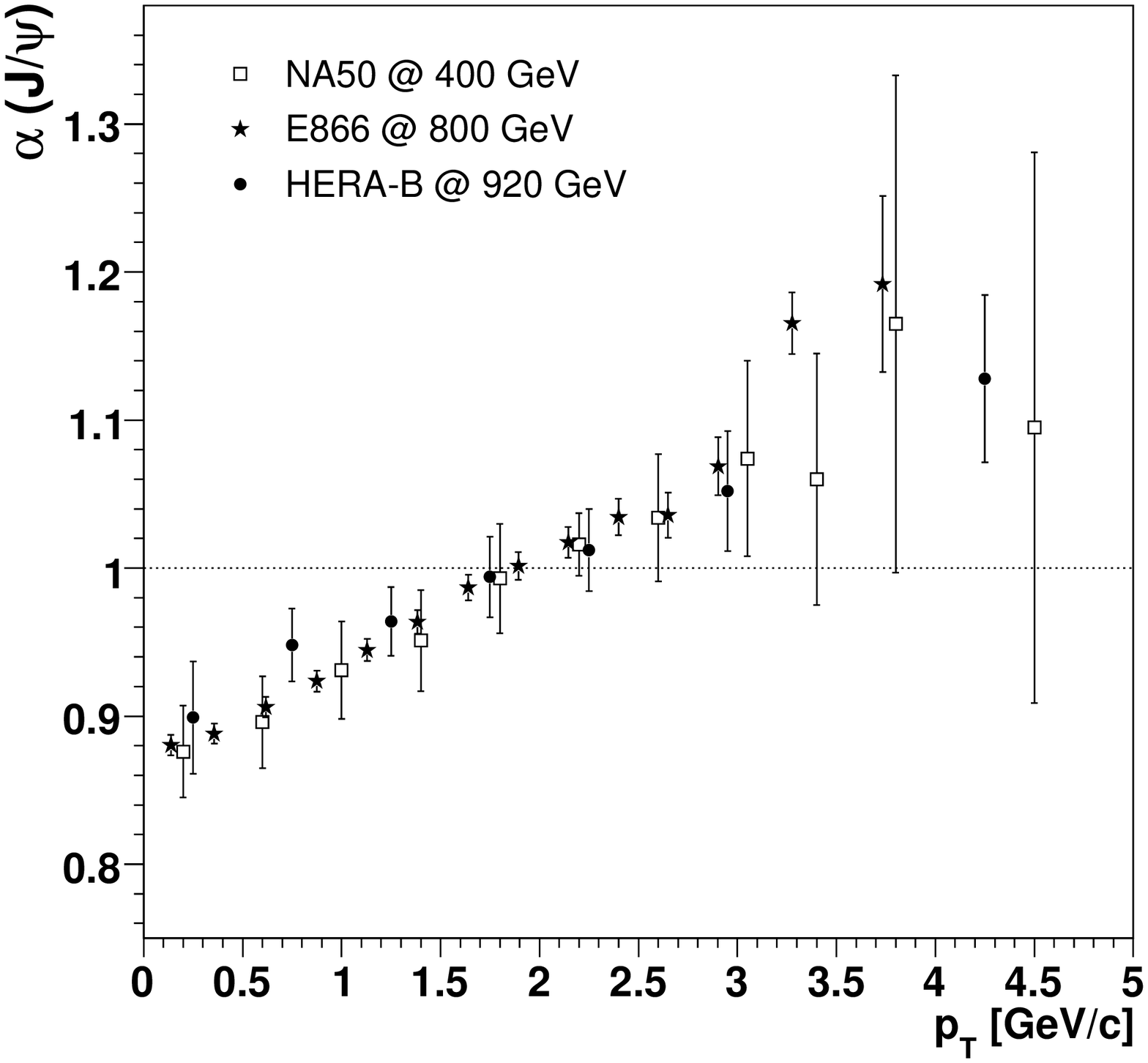}}
\caption{The nuclear-dependence $\alpha$ parameter as a function of
  \xf\ (left) and of \pt\ (right), for \jpsi\ mesons measured at
  three different collision energies, in p-A
  collisions~\cite{alpha-pA}.}
\label{fig:alpha}
\end{figure}

The observation that the \sabs\ values at SPS and RHIC energies are
considerably different raises the question of whether \sabs\ varies
within the energy range covered by the fixed-target experiments.  A
detailed analysis of all the available data, with the Glauber
formalism, including nuclear effects on the PDFs and properly
accounting for the feed-down sources, is beyond the scope of this
paper.  Let us simply state that the (initial and final state) nuclear
absorption of the \jpsi, if parameterized with the $A^\alpha$
expression, leads to values of $\alpha$ that, at mid-rapidity,
increase with collision energy (Fig.~\ref{fig:alpha}-left).

\begin{figure}[h!]
\centering
\resizebox{0.48\textwidth}{!}{%
\includegraphics*{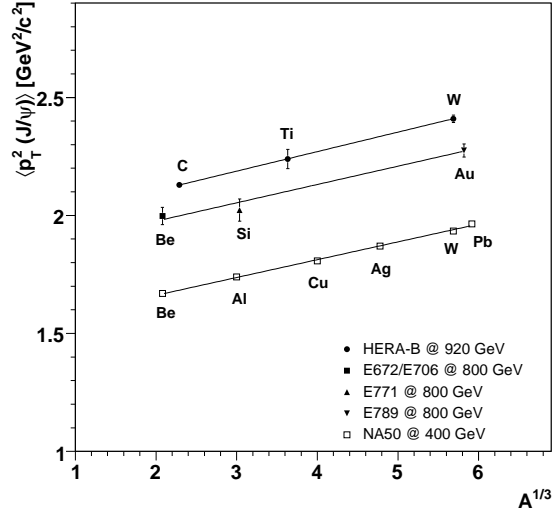}}
\caption{\aptsquared\
  increases with the size of the target nuclei and with the energy of
  the collision~\cite{alpha-pA}.}
\label{fig:pt-vs-A}
\end{figure}  

It is very interesting to notice that the values of $\alpha$ measured
at three significantly different (fixed-target) energies, at
mid-rapidity, show a perfect scaling as a function of \pt\
(Fig.~\ref{fig:alpha}-right).  We may conjecture, then, that the
observed increase of $\alpha$ with energy is actually due to a
convolution between the increase of $\alpha$ with \pt\ and the
increase of the average \pt\ of the \jpsi\ with the collision energy
(Fig.~\ref{fig:pt-vs-A}).

From the previous lines, and also considering that the level of
(anti-)shadowing changes with the collision energy
(Fig.~\ref{fig:npdfs}), it seems natural to deduce that the
\sabs\ value at 158~GeV, the energy of the Pb-Pb collisions, will be
somewhat higher than the value determined at 400/450~GeV.  A stronger
``normal nuclear absorption'' will lead to a decrease of the ``extra
suppression'' presently seen in the Pb-Pb \jpsi\ data.  We will know
more once the NA60 collaboration extracts \sabs\ from the p-A data
collected at 158~GeV.  At this moment, however, in the absence of that
measurement, we cannot really argue that we have a robust reference
baseline at the energy of the heavy-ion data.  Hence, the
``anomalous'' aspect of the \jpsi\ suppression pattern seen in
heavy-ion collisions has not yet been \emph{convincingly} demonstrated, and
requires, in particular, a more detailed understanding of charmonia
production in elementary pp and p-A collisions.

The striking ``change of slope'' of the \psip\ suppression pattern
(Fig.~\ref{fig:na38-na50}-right) looks even more ``anomalous'' but,
again, it occurs between the p-A and the S-U/Pb-Pb data points,
collected with different collision systems and at significantly
different energies.  A convincing demonstration of ``new physics''
would require that the anomaly happens within a consistent set of data
points (ideally, a single collision system, from peripheral to central
collisions).  Unfortunately, poor statistics prevents NA60 from
significantly contributing to the understanding of the \psip\
suppression pattern.  At this moment, and the situation is not likely
to change in the coming years, we cannot say what is the nature of the
``anomaly'' seen in the \psip\ suppression pattern between the p-A and
the S-U/Pb-Pb colliding systems; maybe the extra suppression is due to
the matter produced in the heavy-ion collisions; maybe that matter is
in the QGP phase and ``melts'' the \ccbar\ bound states; maybe this
indicates that the critical energy density is reached already in the
most peripheral S-U or Pb-Pb collisions at the SPS energies;
maybe\ldots

\begin{figure}[ht]
\centering
\resizebox{0.48\textwidth}{!}{%
\includegraphics*{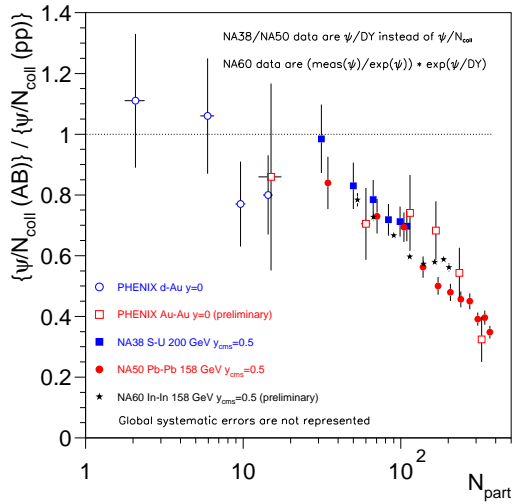}}
\caption{Comparison between the \jpsi\ suppression patterns observed
  at SPS and RHIC energies, as a function of \npart.}
\label{fig:sps-rhic}
\end{figure}

There is much more experimental information concerning the suppression
of the \jpsi\ production yield, with the NA50 Pb-Pb pattern recently
being complemented by PHENIX data taken at 10 times higher centre of
mass energies~\cite{PHENIX-psi} and by NA60 data taken with a smaller
colliding system, In-In~\cite{NA60-psi}.  However, the present
statistical uncertainties of the PHENIX measurements severely limit
the insights gained from comparing the results obtained at both
energies.  Figure~\ref{fig:sps-rhic} is an attempt to integrate
in a single plot, as a function of $N_{\rm part}$ the available
mid-rapidity suppression patterns.  The PHENIX results are $R_{AA}$,
i.e.\ $N^{{\rm J}/\psi}_{\rm Au-Au}(c_i) \,/\, [N^{{\rm
J}/\psi}_{pp}\times N_{\rm coll}(c_i)]$ (and similar for d-Au), where
$c_i$ represents the centrality bin.  The NA38 and NA50 data points
are the same as shown in Fig.~\ref{fig:na38-na50}-left, but normalised
to the pp value, $R^{{\rm J}/\psi}_{AA}(c_i) \,/\, R^{{\rm
J}/\psi}_{\rm pp}$, where $R^{{\rm J}/\psi}$ stands for $B\times
\sigma^{{\rm J}/\psi} \,/\, \sigma^{DY}$.  The NA60 points were
obtained by an approximate procedure, convoluting the In-In ratio
``measured/expected''~\cite{NA60-psi} with the Pb-Pb ``normal nuclear
absorption curve'', $N^{{\rm J}/\psi}(c_i)/G^{{\rm J}/\psi}(c_i)
\,\times\, G^{{\rm J}/\psi\,/\,DY}(c_i) \,/\, G^{{\rm
J}/\psi\,/\,DY}_{pp}$.  In this representation, the different data
sets are directly comparable, since Drell-Yan production scales with
$N_{\rm coll}$.

It is also not simple to compare the SPS heavy-ion patterns.  The S-U
energy is different, it is a very asymmetric collision system, the
Uranium nucleus is not spherical, and, besides, the centrality can
only be determined from the \ET\ measurement, while the In-In
centrality distribution is determined from the \EZDC\ variable.  The
comparison between the In-In and Pb-Pb suppression patterns should
suffer from less systematic uncertainties (same energy, symmetric
collisions, spherical nuclei), especially if we take the Pb-Pb values
from the NA50 \EZDC\ analysis rather than from the \ET\ analysis; but
we must keep in mind that NA50 reports the suppression pattern in
terms of the \jpsidy\ cross-section ratio while NA60 (with much less
high-mass Drell-Yan statistics and a perfect vertexing efficiency even
in the most peripheral collisions) reports the measured \jpsi\ yield
itself, directly normalised by the calculated ``expected'' nuclear
absorption.  These two procedures surely have very different sources
of systematic uncertainties.

\begin{figure}[ht]
\centering
\resizebox{0.48\textwidth}{!}{%
\includegraphics*{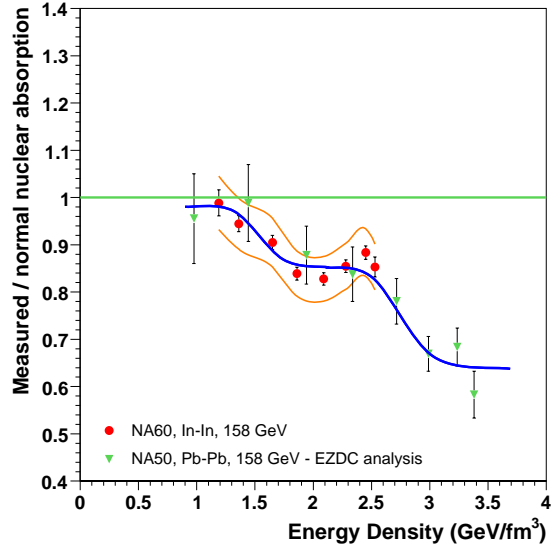}}
\caption{Comparison between the \jpsi\ suppression patterns observed
  in In-In and Pb-Pb collisions at the SPS, as a
  function of the energy density.}
\label{fig:sps}
\end{figure}

Figure~\ref{fig:sps}~\cite{Trento} compares the In-In and
Pb-Pb suppression patterns, as a function of the energy density,
estimated using a calculation based on the VENUS event generator.  The
observed pattern is consistent with a double step function (smeared by
the resolution of the \EZDC\ measurement), as would be expected in
case the \psip\ and \chic\ states would be completely melted in the
medium, above two different thresholds in energy density, leaving only
the (more strongly bound) directly produced \jpsi\ mesons, around
60\,\% of the yield expected in the absence of anomalies~\cite{KKS}.
However, there is a big difference between ``the measurements are
compatible with\ldots'' and ``the measurements show, beyond reasonable
doubt, that\ldots''.

\begin{figure}[ht]
\centering
\resizebox{0.48\textwidth}{!}{%
\includegraphics*{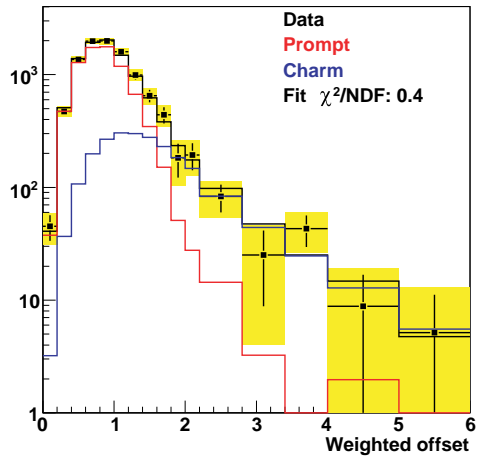}}
\caption{Demonstration that the intermediate mass dimuon excess
  is of prompt origin~\cite{ADavid}.}
\label{fig:imr}
\end{figure}

Concerning the intermediate mass region dimuons, there has been
extremely significant progress provided by the NA60 experiment.  The
analysis reported at this conference~\cite{ADavid} convincingly shows
(thanks to the very good vertexing accuracy of the data) that the
excess dimuons seen in In-In collisions, with respect to the
superposition of Drell-Yan dimuons and simultaneous semi-muonic decays
of D meson pairs, is of a prompt nature (Fig.~\ref{fig:imr}).  

It is worth underlining that this eliminates the long-standing
alternative that the signal excess would, in fact, be due to a small
fraction of un-subtracted background from pion and kaon decays.  It
now remains to be understood what is the physics process responsible
for this source of prompt dimuons.  Maybe we are seeing thermal
dimuons shining from a thermal medium (but, hadronic or deconfined?).
However, once again, in order to clearly disclose any putative new
phenomena, specific of nuclear collisions, we need to understand in
detail the physics processes already contributing in the more
``elementary'' collisions.

\begin{figure}[t]
\centering
\resizebox{0.48\textwidth}{!}{%
\includegraphics*{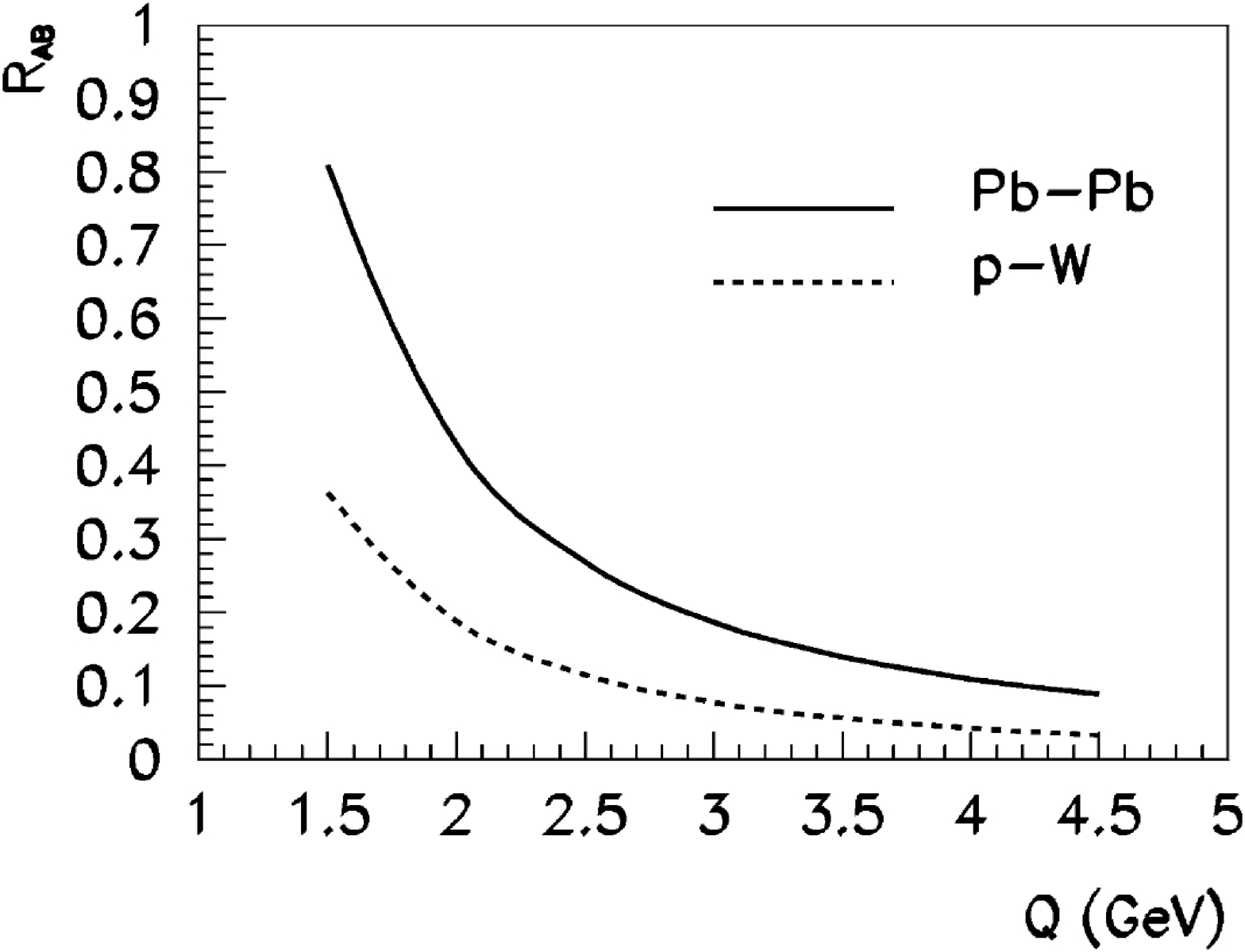}}
\caption{Ratio between the
  double-scattering and the single-scattering contributions to the
  Drell-Yan yield, versus dilepton mass, for p-W and Pb-Pb collisions,
  in the kinematical conditions of the NA60 experiment~\cite{jwqiu}.}
\label{fig:qiu}
\end{figure}

The two physics processes that are expected to give sizeable yields of
dimuons between the $\phi$ and the \jpsi\ peaks are the dimuons from
$q\bar{q}$ annihilation (Drell-Yan) and the simultaneous semi-muonic
decays of correlated charmed hadron pairs.  Unfortunately, the
Drell-Yan yields are not theoretically calculable in a reliable way
for such low values of the dimuon mass, where higher order effects may
be more important than at higher masses.  Furthermore, even if the
perturbative QCD calculations would be consistent with the
proton-proton prompt dimuon continuum down to masses below 1.5~GeV,
this could very well no longer be the case for proton-nucleus and
nucleus-nucleus collisions, where contributions from interactions
involving extra partons from the colliding hadrons may lead to a
significant increase of the low-mass dimuon yield with respect to a
linear scaling of the yield in elementary nucleon-nucleon
interactions~\cite{jwqiu} (Fig.~\ref{fig:qiu}).  Without having
this possible source of \emph{prompt} dimuons under control, through a
detailed analysis of proton-nucleus data, there is little hope that we
can find \emph{clear} evidence for thermal dimuon production in heavy-ion
data.

\begin{figure}[h!]
\centering
\resizebox{0.48\textwidth}{!}{%
\includegraphics*{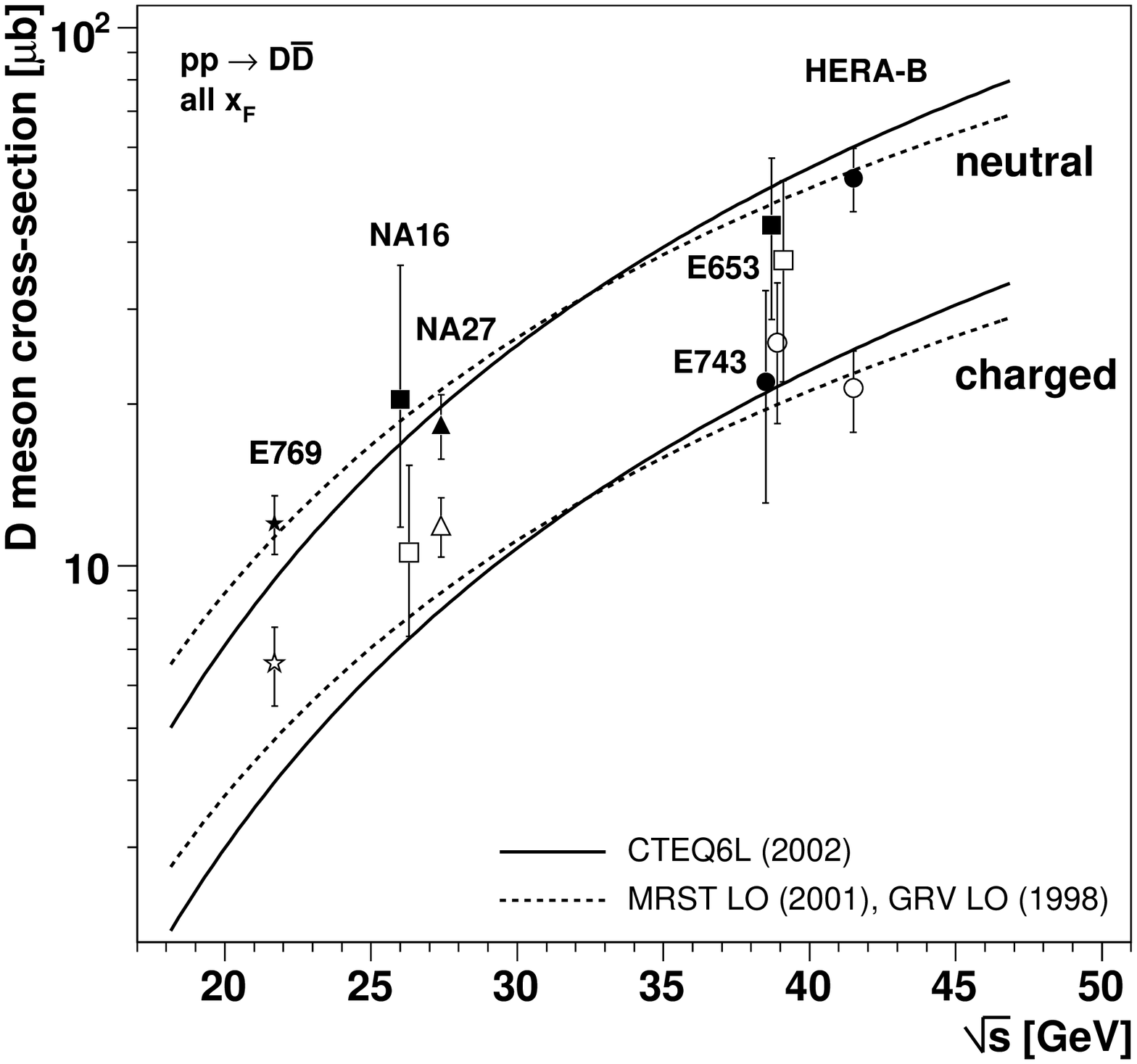}}
\resizebox{0.48\textwidth}{!}{%
\includegraphics*{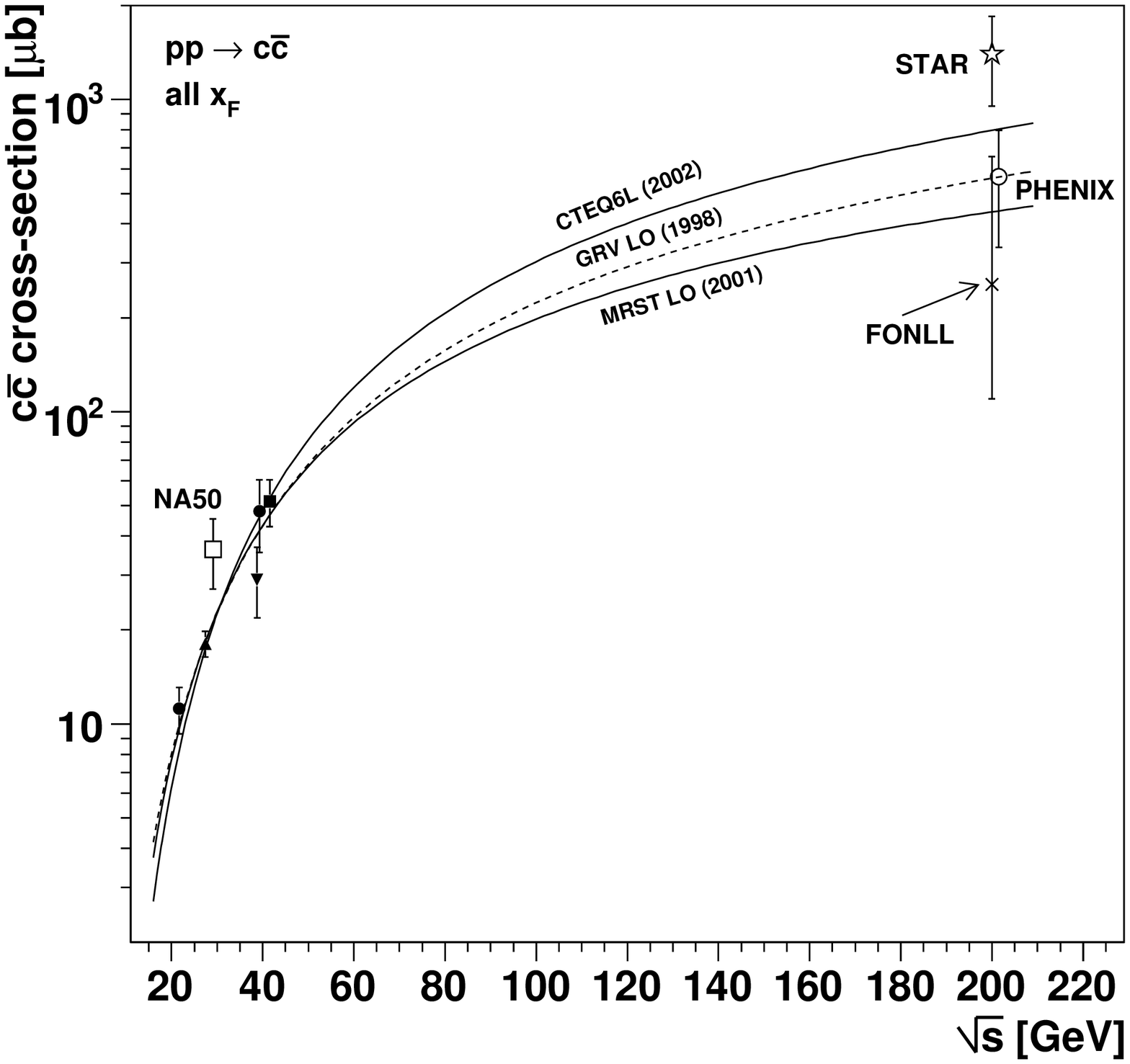}}
\caption{Compilation of production cross sections of charged and
  neutral D mesons (left) and total \ccbar\ (right) as a function of
  \sqrts, compared with curves calculated with Pythia~\cite{Pythia}
  using three different sets of parton densities~\cite{PhysRep}.  The
  value that best describes the NA50 p-A dimuon data~\cite{na50imr} is
  a factor of two higher than expected from the interpolation of the
  direct measurements made by other experiments.  Interestingly, the
  value recently measured by PHENIX~\cite{phenix-charm}, at much
  higher energies, is in good agreement with the extrapolation of the
  fixed-target data.}
\label{fig:charm}
\end{figure}

Also the understanding of the open charm baseline is under question,
given that the charm production cross section deduced from the
measurements made by many experiments (Fig.~\ref{fig:charm}-left) is
two times smaller than the value required to describe the dimuon data
collected by NA50 in p-A collisions (Fig.~\ref{fig:charm}-right).
This indicates that there seems to be something ``anomalous'' already
in the intermediate mass dimuons produced in the p-A collisions
studied by NA50.

Reliable baseline measurements, made in elementary collisions, are
also indispensable to correctly interpret photon and low mass dilepton
production.  In both cases, a large background from hadron decays
needs to be removed so that we access a rare signal, which then needs
to be understood as a convolution of several physics sources, maybe
including thermal radiation from the QGP and/or from a hot hadron
gas~\cite{GDavid}.

In the case of low mass dilepton production, the exceptional increase
in statistics between the CERES Pb-Au data and the NA60 In-In data
implies that the significance of the measurements is now determined by
systematic uncertainties, related to experimental aspects like
efficiencies and acceptances (increasingly difficult to keep under
control as we move down in mass and \pt) and to the definition of the
hadronic decay ``cocktail'', given our limited knowledge of
kinematical distributions, $\rho$/$\omega$ interference effects, the
$\omega$ dimuon branching ratio, the $\omega$ form factor, the $\rho$
mass line shape, and other ``inputs'' needed to define the ``expected
sources'' reference.

We must also keep in mind that there are ``normal nuclear effects'' at
play in low mass dilepton production, such as the increase of the
$\phi/\omega$ cross-section ratio in p-A collisions~\cite{HKW-HP04}.
How much of the $\phi$ enhancement seen in heavy-ion collisions
remains ``anomalous'' after accounting for a correct extrapolation of
the p-A observations?  And how can we conciliate our basic
understanding of low mass dilepton production in ``elementary''
collisions with E325's claim~\cite{E325} that there are already
``nuclear matter modifications'' of the properties of the $\rho$,
$\omega$ and $\phi$ mesons produced in p-Cu collisions at 12~GeV?

In summary, while it is certainly true that there has been very
considerable progress in the recent years regarding the diversity and
quality of the measurements related to quarkonium and electromagnetic
probes production in high-energy heavy-ion collisions, it remains a
field with many open questions.  Some of them might be given
satisfactory answers once the data samples recently collected at the
SPS and at RHIC are fully analysed.  Others will presumably remain
unanswered and would easily justify significant upgrades of existing
experiments or even the construction of new ones, provided the
allocated integrated beam time (and human resources) would properly
match the investment required to build the new hardware.  In any case,
the present situation, after 20 years of efforts by many hundreds of
physicists, remains unsatisfactory: we do not yet have convincing
evidence, from experimental data on quarkonium or electromagnetic
probes, that shows \emph{beyond reasonable doubt} the existence of ``new
physics'' in high-energy heavy-ion collisions.

\bigskip

This paper benefited from valuable discussions with several people,
including, in particular, 
David d'Enterria,
Gon\c{c}alo Borges,
Helena Santos,
Helmut Satz,
Hermine W\"ohri,
Louis Kluberg,
Mike Leitch,
Pietro Faccioli,
Ramona Vogt,
Rapha\"el Granier de Cassagnac
and Roberta Arnaldi.

\end{document}